\documentclass{ifacconf}

\usepackage{natbib}        
\usepackage{graphicx} 
\usepackage{ mathtools, amssymb, mathrsfs}  
\usepackage{enumerate} 
\usepackage{array,multirow}
\usepackage[table]{xcolor}
\usepackage{ diagbox}
\usepackage{ empheq}

\newtheorem{theo}{Theorem}

\newtheorem{lemma}{Lemma}
\newtheorem{property}{Property}
\begin{document}
\begin{frontmatter}

\title{Connections Between Instrumental Variable, Pseudo-Linear Regression, and Prediction Error Methods in Terms of Their Minimized Criteria} 

\author[Vau]{Bernard Vau} 
\author[Bourles]{Henri Bourl\`es}

\address[Vau]{EXAIL, 
   94380 Bonneuil-sur-Marne, France (e-mail: bernard.vau@exail.com)}
\address[Bourles]{CNAM, SATIE, France (e-mail: henri.bourles@lecnam.net}

\begin{abstract}                
This paper explores the connections between the criteria directly or indirectly minimized by Prediction Error (PE), Instrumental Variable (IV), and Pseudo-Linear Regression (PLR) algorithms for the identification of output-error models. By definition, PE methods aim to minimize the variance of the prediction error. In contrast, the correlation functions appearing in the stationarity conditions of IV and PLR algorithms are not the gradients of any underlying loss function involving a signal variance. Nevertheless, if the true system lies in the model set, we show that one IV algorithm minimizes asymptotically the variance of an unmeasurable signal called the equivalent prediction error. This signal was introduced previously to analyze the bias distribution of PLR schemes. It turns out that, regardless of whether the true system lies in the model set, some iterative IV and PLR algorithms aim to minimize the prediction error variance and can therefore be interpreted as indirect PE algorithms.
\end{abstract}


\end{frontmatter}

\section{Introduction}
Classical parametric identification methods continue to play a central role in system identification, motivating ongoing efforts to refine their theoretical analysis and better understand their properties. According to (\cite{r5}), the three classical approaches to the parametric identification of discrete-time Linear Time Invariant (LTI) systems represented by transfer functions are Prediction Error Methods (PEM), Instrumental Variable (IV) methods, and Pseudo-Linear Regression (PLR) methods. PEM explicitly minimize the variance of the prediction error between the system output and the predicted model output. The IV method with an auxiliary model, called SIV in (\cite{r13}) (see also (\cite{r6}), seeks to achieve zero correlation between an equation error and a vector composed of past inputs and outputs of a simulated model, which are independent of the disturbing noise. On the other hand, the PLR Output Error scheme (called Landau's scheme in (\cite{r5}), or OLOE) aims to achieve zero correlation between an output error and a vector also composed of inputs and outputs of a simulated model. PEM and IV are currently the dominant approaches, whereas PLR schemes are still employed in the field of recursive identification (\cite{r3}), especially for indirect adaptive control purposes. IV and PLR belong to what is referred to in (\cite{r5}) as the correlation approach. Research on IV methods in system identification, and on the correlation approach in general, remains an active area of investigation, (\cite{r29}), (\cite{r28}), (\cite{r30}), (\cite{r32}), (\cite{r31}). \\
\noindent This paper focuses on the identification of output-error models and establishes connections between the criteria directly or indirectly minimized by PEM, IV, and PLR. Whereas a criterion minimized by IV schemes is known and expressed from generalized moments  (\cite{r27}), (\cite{r6}), it is useless for an interpretation of the bias distribution, since this criterion is not related to the minimization of a signal variance that could be used to analyze the asymptotic bias distribution in the frequency domain, as was done many years ago for PEM (\cite{r25}). Moreover, the correlation function appearing in the stationarity conditions of IV (and also of PLR) schemes is not the gradient of any putative loss function  This does not mean, however, that SIV and OLOE do not indirectly minimize the variance of a signal. \\
\noindent In order to study the minimized criterion of certain PLR schemes, the concept of equivalent prediction error was introduced in (\cite{r4}). This is an unmeasurable signal whose variance is effectively minimized by PLR algorithms when the true system belongs to the model set. It was also shown in (\cite{r22}) that this property remains approximately valid (but still relevant) when the true system does not belong to the model set, provided that the bias is small. Moreover, it was proved that the criterion minimized by PLR schemes depends on the predictor parameterization. This led to the development of novel PLR schemes using models parameterized in terms of Generalized Orthonormal Basis Functions (GOBFs), in (\cite{r2}) for closed-loop algorithms and in (\cite{r23}) for the open-loop output-error case. The latter scheme is called H-OLOE in (\cite{r23}), where an iterative procedure, called I-H-OLOE, was also proposed to achieve statistical efficiency through successive H-OLOE runs. Furthermore, a link was made in (\cite{r22}) between the asymptotic bias distribution of SIV and OLOE, using the concept of equivalent prediction error:  in the presence of bias, the limit models obtained by these two schemes both minimize the same criterion. \\
\noindent The purpose of this paper is to further investigate the connections between IV and PLR output-error schemes when the system input-output data are filtered in the IV scheme and when the PLR algorithm is H-OLOE. It turns out that the SIV scheme using data filtered by an autoregressive (AR) filter minimizes the same criterion as H-OLOE when its basis poles coincide with the poles of this AR filter. Finally, it is shown that the iterative IV scheme called SRIV and the iterative PLR scheme I-H-OLOE, if they converge, both asymptotically minimize a criterion that is precisely the output-error variance, regardless of whether the true system lies in the model set. This finding sheds new light on these schemes, since they can be interpreted as indirect ways of solving a PEM problem. \\
\noindent The paper is organized as follows. Section II provides an overview of the stationarity conditions of PEM, IV, and PLR and establishes a connection between the latter two classes of methods. Section III presents the common criterion minimized by SIV and H-OLOE. Section IV analyzes the effects of data filtering and predictor parameterization on IV and PLR schemes. Finally, Section V shows that both SRIV and I-H-OLOE aim to minimize the prediction error variance and can therefore be interpreted as indirect ways of solving the output-error PEM problem.

\section{Review of the stationary conditions of IV, PLR and PEM schemes}

Let us consider a "true" system $G_0(q, \theta_0)$ to be identified in open-loop, excited by an input sequence $\{u(t)\}$. One assumes that the system is corrupted by a centered output error noise sequence $\{v(t)\}$, such that the output sequence $\{y(t)\}$ can be expressed as
\begin{equation}
{y(t)=G_0(q,\theta_0)u(t)+v(t)=\frac{B_0(q,\theta_0)}{A_0(q,\theta_0)}u(t)+v(t)} \\
\label{e1}
\end{equation}
 \noindent In \eqref{e1} $q$ is the forward shift operator, $\theta_0$ the true system parameter vector. One seeks to estimate $\theta_0$ by means of a model $G(q, \theta)$ parameterized with the vector $\theta$ such that
  \begin{equation}
G(q,\theta)=\frac{B(q,\theta)}{A(q,\theta)}=\frac{b_1q^{n_a-1}+b_2q^{n_a-2}+\cdots +b_{n_a}}{q^{n_a}+a_1q^{n_a-1}+a_2q^{n_a-2}+\cdots +a_{n_a}}
\label{e1b}
\end{equation}
 \noindent One defines the output error $\varepsilon_{\scalebox{0.6}{$OE$}}(t+1,\theta)=y(t+1)-\hat{y}_{\scalebox{0.6}{$OE$}}(t+1,\theta)$ where $\hat{y}_{\scalebox{0.6}{$OE$}}(t,\theta)=G(q,\theta)u(t)$. The prediction error of the output error model is
 \begin{equation}
 \varepsilon_{\scalebox{0.6}{$OE$}}(t,\theta)=\left(G_0(q,\theta_0)-G(q,\theta)\right)u(t)+v(t)=y(t)-\hat{y}_{\scalebox{0.6}{$OE$}}(t,\theta)
\label{eq1c}
 \end{equation}
 \noindent This predicted output can be expressed under a pseudo-linear regression form
 \begin{equation}
 \hat{y}_{\scalebox{0.6}{$OE$}}(t+1,\theta)=\theta^T\phi(t,\theta)
 \label{eq1d}
 \end{equation}
  \noindent with 
 \begin{subequations}
 \begin{align}
{  \phi^T(t)=\left[ -\hat{y}_{\scalebox{0.6}{$OE$}}(t,\theta) \; -\hat{y}_{\scalebox{0.6}{$OE$}}(t-1,\theta) \cdots  u(t) \; u(t-1) \cdots  \right]  } \\
 { \theta^T=\left[a_1 \; a_2 \cdots  b_1 \; b_2 \right]}
  \end{align}
  \label{eq_phi}
  \end{subequations}
  \noindent One also defines an equation error $\varepsilon_{LS}(t+1,\theta)=y(t+1)-\hat{y}_{\scalebox{0.6}{$LS$}}(t+1,\theta)$ where 
 \begin{equation}
 {\hat{y}_{\scalebox{0.6}{$LS$}}(t+1,\theta)=\theta^T\phi_{LS}(t,\theta)}
  \label{eq1e}
 \end{equation}
   \noindent and
  \[ \phi_{LS}^T(t)=\left[ -y(t) \; -y(t-1) \cdots  u(t) \; u(t-1) \cdots  \right] \]
   \noindent and one verifies easily that 
   \begin{multline}
  {\varepsilon_{LS}(t+1,\theta)=A(q,\theta)\varepsilon_{\scalebox{0.6}{$OE$}}(t,\theta)} \\
  {=A(q,\theta)\left(G_0(q)-G(q,\theta)\right)u(t)+A(q,\theta)v(t)}
  \label{eq1f}
  \end{multline}
  \noindent One assumes that all the signals are quasi-stationary (see (\cite{r5}), and one introduces the following notation ($\mathbf{E}[x(\tau)]$ being the expectation of $x(\tau)$)\\
 \[ \bar{\mathbf{E}}[x]=lim_{N \to \infty}   \frac{1}{N+1}\sum_{\tau=0}^{N}\mathbf{E}[x(\tau)]                     \] 
 \noindent The classical Least-Squares (LS) algorithm consists in minimizing the criterion $J_{LS}=\frac{1}{N}\sum_{t=0}^{N-1}\varepsilon_{LS}^2(t+1,\theta)$. Therefore the asymptotic estimated parameters vector $\theta^{*}_{LS}$ satisfies asymptotically the stationary condition
 \begin{equation}
 \bar{\mathbf{E}}[\varepsilon_{LS}(t+1,\theta^*_{LS})\phi_{LS}(t)]=0
 \label{eq_stationar_LS}
 \end{equation}
 \noindent and it is well known that, except if the true system has an ARX structure, the LS estimator is biased, because of the correlation between $\phi_{LS}(t)$ and $v(t+1)$. In order to overcome this issue, two very popular identification methods were designed decades ago: \\
 \noindent 1) The Instrumental variable method with auxiliary method called SIV in (\cite{r13}), (see also (\cite{r6}), where the asymptotic estimated vector $\theta^{*}_{SIV}$ satisfies the stationary condition
 \begin{equation}
 \bar{\mathbf{E}}[\varepsilon_{LS}(t+1,\theta^*_{SIV})\phi(t,\theta^*_{SIV})]=0
 \label{eq_stationar_SIV}
 \end{equation}
 \noindent The estimator consistency being obtained since $\phi(t,\theta)$ (called the instrument) is, by construction, uncorrelated with $v(t+1)$. The estimator can be estimated recursively or off-line (see (\cite{r13}) p.173). \\
 \noindent 2) The Pseudo Linear Regression Output Error scheme (called the Landau's scheme in (\cite{r5}), or OLOE (see chap. 3 of (\cite{r3}) whose stationary condition entails ($\theta^*_{PLR}$ being the asymptotic estimated parameters vector)  
 \begin{equation}
 \bar{\mathbf{E}}[\varepsilon_{\scalebox{0.6}{$OE$}}(t+1,\theta^*_{PLR})\phi(t,\theta^*_{PLR})]=0
 \label{eq_stationar_PLR}
 \end{equation}
 \noindent In PLR schemes, the estimated parameters vector is most of the time computed recursively with the so-called Parameter Adaptation/Learning Algorithm (PALA), see Chap.3 and 4 of (\cite{r3}). 
 \noindent The estimation algorithms of SIV are not recalled since they have now become very classical. The stationary condition of OLOE \eqref{eq_stationar_PLR} differs from the stationary condition of the output error of PEM, since this last method aims directly at minimizing the criterion $J_{PEM}$ with ($\theta^*_{PEM}$ the estimated parameters vector of PEM)
 \begin{equation}
  J_{PEM}=\bar{\mathbf{E}}[\varepsilon^2_{OE}(t,\theta)]
  \label{eq_critere_PEM}
  \end{equation}
   and its stationary condition is therefore 
 \begin{equation}
 \bar{\mathbf{E}}\left[\varepsilon_{\scalebox{0.6}{$OE$}}(t+1,\theta^*_{PEM})\left. \frac{\partial \varepsilon_{\scalebox{0.6}{$OE$}}(t+1,\theta)}{\partial \theta} \right|_{\theta=\theta^*_{PEM}} \right]=0 
 \label{eq_stationar_PEM}
 \end{equation}
 \noindent Contrary to PEM, the vector fields associated to the stationary condition of SIV and OLOE schemes are not the gradient of any loss function, since it can be easily verified that their corresponding Jacobian matrices are not symmetric in general (which is contrary to the condition of  the Poincarré Lemma -see Theorem 5-4 of (\cite{r21}), for a vector flow be the gradient of a function). However this does not mean that these two algorithms indirectly minimize any loss function, as shown in the sequel.

 \section{The criterion minimized by SIV and OLOE schemes}

 \subsection{Review of the theory of the equivalent prediction error}
 
 Let us define, as in (\cite{r4}) a signal denoted as $\varepsilon_E(t,\theta)$ that we call the equivalent prediction error, and a vector called the equivalent regressor denoted as $\phi_E(t)$ independent of $\theta$ such that 
\begin {equation}
\phi_{E}(t)=-\frac{\partial \varepsilon_E(t+1,\theta)}{\partial \theta}
\label{eq_eq_regressor}
\end{equation}
\noindent Let us assume the following properties
\begin{subequations}
\begin{align}
{\varepsilon_E(t,\theta_0)=\varepsilon_{\scalebox{0.6}{OE}}(t,\theta_0)=v(t)} \\
{\phi_E(t)=\phi(t,\theta_0)}
\end{align}
\label{eq_hyp_eq_pred}
\end{subequations}
\noindent PLR estimators are known to be consistent if the true system is in the model set, and assume temporarily that this holds true, then from \eqref{eq_stationar_PLR} 
\begin{equation}
 \bar{\mathbf{E}}[\varepsilon_{E}(t+1,\theta_0)\phi_E(t)]=0
 \label{eq_stationar2_PLR_th0}
 \end{equation}
\noindent and owing to the expression of $\phi_E(t)$ in \eqref{eq_eq_regressor}, the criterion minimized by PLR schemes $J_{PLR}$ satisfies
\begin{equation}
J_{PLR}=\bar{\mathbf{E}}[\varepsilon_{E}^2(\theta)]
\label{eq_critere_PLR}
\end{equation}
\noindent The expression of the equivalent prediction error was first stated in (\cite{r4}). We provide here a more rigorous derivation of this expression:

\begin{theo}
Properties \eqref{eq_hyp_eq_pred} concerning $\varepsilon_E(t,\theta)$ and $\phi_E(t)$ hold if and only if 
\begin{equation}
\varepsilon_{E}(t+1,\theta)=Q(q,\theta)\varepsilon_{\scalebox{0.6}{OE}}(t+1,\theta)+(1-Q(q,\theta))v(t+1)
\label{eq_eq_pred_err}
\end{equation}
\noindent  where the transfer function operator $Q(q,\theta)$, specific for each PLR scheme is here
\begin{equation}
 \phi(t,\theta)=-Q(q,\theta)\frac{\partial\varepsilon_{\scalebox{0.6}{OE}}(t+1,\theta))}{\partial\theta}
 \label{eq_Q2}
 \end{equation}
 \label{theo_eq_eq_pred}
\end{theo}
\noindent Proof: 1) Sufficiency part: One verifies that taking \eqref{eq1c} and \eqref{eq_eq_pred_err} at $\theta=\theta_0$ yields $\varepsilon_E(t,\theta_0)=\varepsilon_{\scalebox{0.6}{OE}}(t,\theta_0)=v(t)$. 
Moreover $\frac{\partial\varepsilon_E(t+1,\theta)}{\partial\theta}=\frac{\partial Q(q,\theta)}{\partial\theta}(\varepsilon_{\scalebox{0.6}{OE}}(t+1,\theta)-v(t+1))+Q(q,\theta)\frac{\partial \varepsilon_{\scalebox{0.6}{OE}}(t+1,\theta)}{\partial \theta}$. Taking  $\theta=\theta_0$ and combining with \eqref{eq_Q2} entails $\phi_E(t)=\phi(t,\theta_0)$, owing to the definition \eqref{eq_eq_regressor}. \\
\noindent 2) 
Necessity part: By deriving \eqref{eq1c}, \eqref{eq1d}, one gets\footnote{The use of partial derivatives for signals defined on extended Hilbert spaces is justified in Appendix of \cite{r4}. Set $\{x(t)\},\{w(t)\} \in l_{2e}$  where  $l_{2e}$ denote the extended $l_2$ space consisting of all sequences that are square summable over a finite horizon. If $w(t)=\mathcal{G}(q)x(t)$, one has $\frac{\partial w}{\partial x}=\mathcal{G}(q)$.} 
$\frac{\partial q\varepsilon_{\scriptscriptstyle OE}}{\partial \theta}=-\phi-\theta^T\frac{\partial{\phi}}{\partial q\varepsilon_{\scriptscriptstyle OE}}\frac{\partial q\varepsilon_{\scriptscriptstyle OE}}{\partial \theta}$. 
Thus from \eqref{eq_Q2}, $Q=1+\theta^T\frac{\partial\phi}{\partial q\varepsilon_{\scriptscriptstyle OE}}$, and one has $\frac{\partial Q}{\partial \theta}=\frac{\partial\phi}{\partial q\varepsilon_{\scriptscriptstyle OE}}$ (since  $\frac{\partial}{\partial \theta}\left(\frac{\partial \phi}{\partial q\varepsilon_{\scriptscriptstyle OE}}\right)=0$). \\
One the other hand, if the true system lies in the model set from \eqref{eq1d}, $\hat{y}_{\scalebox{0.6}{$OE$}}(t,\theta_0)=G_0(q,\theta_0)u(t)=y(t)-v(t)$, and owing to \eqref{eq_phi}, the entries of $\phi_E(t)=\phi(t,\theta_0)$ are the delayed noise-free outputs or the delayed inputs of $G_0$, and one observes that $\phi(t,\theta)-\phi_E(t)=\frac{\partial\phi}{\partial q\varepsilon_{\scriptscriptstyle OE}}(q\varepsilon_{\scriptscriptstyle OE}-qv)$. From \eqref{eq_Q2} $-\phi_E(t)=\frac{\partial\phi}{\partial q\varepsilon_{\scriptscriptstyle OE}}(q\varepsilon_{\scriptscriptstyle OE}-qv)+Q\frac{\partial (q\varepsilon_{\scriptscriptstyle OE}-qv)}{\partial \theta}=\frac{\partial Q}{\partial \theta}(q\varepsilon_{\scriptscriptstyle OE}-qv)+Q\frac{\partial (q\varepsilon_{\scriptscriptstyle OE}-qv)}{\partial \theta}$. Now integrating with respect to $\theta$ entails $q\varepsilon_E=Q(q\varepsilon_{\scriptscriptstyle OE}-qv)+k$. Taking $\theta=\theta_0$ yields $k=qv$, which is \eqref{eq_eq_pred_err}. $\Box$


\noindent The expression of the criterion $J_{PLR}$ is rigorously exact if the true system is in the model set and only then\footnote{Contrary to what was implicitly claimed in (\cite{r4}). The authors express their gratitude to Professor Xavier Bombois who pointed out this issue.}. It was shown in (\cite{r22}) that this becomes approximate but still relevant provided that the bias is small in a given sense precisely defined in this reference. 
 \noindent The following result holds:
 
 \begin{lemma}
 The PLR stationary condition \eqref{eq_stationar_PLR} is equivalent to 
 \begin{equation}
\bar{\mathbf{E}}[\varepsilon_{E}(t+1,\theta^*_{PLR})\phi(t,\theta^*_{PLR})]=0
\label{eq_stationar3_PLR}
\end{equation}
\noindent regardless of whether the true system lies in the model set.
\end{lemma}
\noindent Proof: The proof follows from Theorem 1 in (\cite{r22}). $\Box$  

\noindent From \eqref{eq_Q2}, another relation can be stated:
\begin{property}
\label{prop00}
The prediction error satisfies
\begin{equation}
\varepsilon_{\scalebox{0.6}{OE}}(t+1,\theta))=Q^{-1}(q,\theta_0)\left(\theta_0-\theta\right)\phi(t,\theta)+v(t+1)
\label{eq_pred_QO}
\end{equation}
\end{property}
\noindent Proof: From \eqref{eq1d} $\frac{\partial q\varepsilon_{\scalebox{0.6}{OE}}}{\partial \theta}=-\phi-\theta^T\frac{\partial\phi}{\partial \theta}=-\phi+(\theta_0-\theta)^T\frac{\partial\phi}{\partial \theta}-\theta_0^T\frac{\partial \phi}{\partial q\varepsilon_{\scalebox{0.6}{OE}}}\frac{\partial q\varepsilon_{\scalebox{0.6}{OE}}}{\partial \theta}$. Thus (see proof of Theorem \ref{theo_eq_eq_pred}) $Q(\theta_0)\frac{\partial q\varepsilon_{\scalebox{0.6}{OE}}}{\partial \theta}=-\phi+(\theta_0-\theta)^T\frac{\partial\phi}{\partial \theta}$. Integrating with respect to $\theta_0-\theta$ ($\theta_0$ is constant) yields $Q(\theta_0)q\varepsilon=(\theta_0-\theta)^T\phi+k$. Taking $\theta=\theta_0$ yields $k=Q(\theta_0)qv$, hence the result. $\Box$

\noindent The relation \eqref{eq_pred_QO} has been determined by another means in (\cite{r3} chap. 3 and 4), and is widely used for PLR schemes stability analysis. 

\subsection{Criteria of OLOE and SIV}
 
\noindent Now, combining   \eqref{eq1c}, \eqref{eq1d}, \eqref{eq_phi}, and \eqref{eq_Q2}  yields for the OLOE scheme
\begin{equation}
Q(q,\theta)=A(q,\theta)
\label{eq_Q_OLOE1}
\end{equation}
\noindent thus, from \eqref{eq_eq_pred_err} the explicit expression of the equivalent prediction error of OLOE is
\begin{equation}
\varepsilon_{E}(t+1,\theta)=A(q,\theta)\varepsilon_{\scalebox{0.6}{$OE$}}(t+1,\theta)+(1-A(q,\theta))v(t+1)
\label{eq_eq_pred_err_OLOE}
\end{equation}
\noindent similarly
\begin{equation}
\varepsilon_{E}(t,\theta)=A(q,\theta)\left(G_0(q)-G(q,\theta)\right)u(t)+v(t)
\label{eq_eq_pred_err_OLOE_direct}
\end{equation}
\noindent but combining \eqref{eq_eq_pred_err_OLOE_direct}   with \eqref{eq1f} yields another expression of this equivalent prediction error
 \begin{equation}
\varepsilon_{E}(t+1,\theta)=\varepsilon_{LS}(t+1,\theta)+(1-A(q,\theta))v(t+1)
\label{eq_eq_pred_err2}
\end{equation}
\noindent and finally combining  \eqref{eq_stationar3_PLR} of Lemma 1 with  \eqref{eq_eq_pred_err2} entails (since $\phi(t, \theta)$ is independent of the noise $v(t+1)$ by assumption)
\begin{equation}
\bar{\mathbf{E}}[\varepsilon_{LS}(t+1,\theta^*_{\scalebox{0.6}{$OLOE$}})\phi(t,\theta^*_{\scalebox{0.6}{$OLOE$}})]=0
\label{eq_stationar4_PLR}
\end{equation}
\noindent where $\theta^*_{\scalebox{0.6}{$OLOE$}}$ is the asymptotic estimate of OLOE. This is nothing other than the stationary condition of the SIV algorithm given in \eqref{eq_stationar_SIV}, and one concludes that even if the true system is not in the model set one has 
\begin{equation}
\theta^*_{SIV}=\theta^*_{\scalebox{0.6}{$OLOE$}}
\label{eq_SIV_PLR}
\end{equation}
\noindent showing that the limit models of SIV and OLOE are exactly the same, even if the true system does not lie in the model set. Therefore, considering \eqref{eq_eq_pred_err_OLOE_direct} and the criterion $J_{PLR}$ given in \eqref{eq_critere_PLR}, by shifting in the frequency domain, the following result holds: 

\begin{property}
Consider the asymptotic parameters vectors $\theta^*_{SIV}$ of SIV and $\theta^*_{\scalebox{0.6}{$OLOE$}}$ of OLOE. Assume that the identified models both have the same order $n_a$. Then the following equality holds
 \[\theta^*_{SIV}=\theta^*_{\scalebox{0.6}{$OLOE$}}  \]
 \noindent regardless of whether the true system lies in the model set. Moreover, if the true system lies in the model set, the criterion minimized by both schemes is $J_{1}$ defined by ($\Phi_{u}(\omega)$ being the Power Spectrum density (PSD) of $\{u(t)\}$)
\begin{multline}
{ J_{1}= } \\
{\frac{1}{2\pi}    \int_{-\pi}^{\pi} \left|{A}(e^{i\omega},\theta)\right|^2  \lvert G_0(e^{i\omega})-{G}(e^{i\omega},\theta) \rvert^{2}\Phi_{u}(\omega)      d{{\omega}} }
\label{limit_mod_PLR_SIV}
\end{multline}
\label{prop1}
\end{property}
 
 \noindent  The criterion \eqref{limit_mod_PLR_SIV} remains almost exact if the true system no longer lies in the model set, provided that the bias is small in a given sense precisely stated in (\cite{r22}). One can compare this result to the limit model of the Output Error PEM scheme: combining \eqref{eq_critere_PEM} and \eqref{eq1c}, and shifting in the frequency domain yields
 \begin{multline}
{\theta^*_{PEM}=  } \\
{  \operatorname*{arg\,min}_{\theta} \frac{1}{2\pi}  \int_{-\pi}^{\pi}   \lvert G_0(e^{i\omega})-{G}(e^{i\omega},\theta) \rvert^{2}\Phi_{u}(\omega)      d{{\omega}} }
\label{limit_mod_PEM}
\end{multline}
\noindent The expressions \eqref{limit_mod_PLR_SIV} and \eqref{limit_mod_PEM} show that SIV and PLR schemes heavily penalize the misfit at high frequencies due to the weighting term $\left|{A}(e^{i\omega})\right|^2$. In general, this property is not advantageous, which motivates the modification of this criterion through an appropriate strategy.

\subsection{Simulation example}
\begin{figure}
\begin{center}
\includegraphics[ width=3.4in, keepaspectratio ]{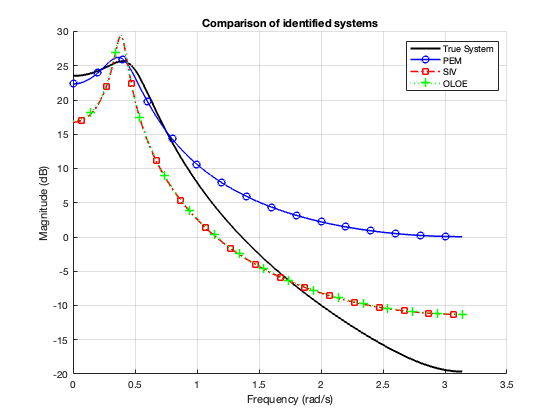}
\caption{Comparison of the limit models of SIV, OLOE and PEM}
\label{fig1}
\end{center}
\end{figure}

Consider the system
\[
G_0(q)=\frac{q^{-2}+0.5q^{-3}}{(1-1.5q^{-1}+0.7q^{-2})(1-0.5q^{-1})}
\]
with additive output noise $v(t)$, assumed to be white and zero-mean, and a noise-to-output variance ratio of $-10$ dB. The system, excited by a white input sequence, is identified in open-loop using SIV and OLOE, with $n_a=2$, so that the true system does not belong to the model set. The identification is performed using a very large number of samples ($N=1000000$) to obtain models as close as possible to their asymptotic limits. The identified models are compared with the model obtained using Output Error PEM. It is clear that the models obtained using SIV and OLOE are identical, as expected, and differ from the model obtained using PEM (see Fig.~\ref{fig1}). The SIV and OLOE models provide a better fit at high frequencies, whereas PEM provides a better fit at low frequencies.


\section{Effect of data filtering and of predictor parametrization}

\subsection{Expression of the criterion}

A novel parameterization based on GBOFs was proposed in (\cite{r2}) for models identified using PLR schemes (output-error and ARMAX models in open- or closed-loop operation). In an open-loop context, the output-error algorithm is called H-OLOE in (\cite{r23}) and (\cite{r22}). Its main advantage is its ability to robustly identify fast-sampled systems and processes with modes separated by several decades (stiff systems). 
GBOFs are derived from a balanced realization of an all-pass function $G_{b}(z)=\prod_{j=0}^{n_{p}-1}\frac{-z.p_j+1}{z-p_{j}}$, where $p_j$ are the basis poles (strictly inside the unit circle) and $n_p$ is the number of basis poles. There exists a balanced state-space realization such that $G_{b}(z)=D_b+C_{b}\left(zI-A_b\right)^{-1}B_b$ (for the construction of the state-space matrices, see Chapter 2 of (\cite{r1}). The orthonormal transfer function vectors $V_k(z)$, with $k=1,2,\cdots$ and dimensions $(n_p,1)$, are given by $V_{k}(z)=\left(zI-A_b\right)^{-1}B_{b}G_{b}^{k-1}(z)$. Orthonormality of these functions follows from the balanced state-space realization of $G_b(z)$. Particular choices of $n_p$ and $p_j$ correspond to well-known cases: $n_p=1, p_{0}=0$ yields the classical $z^{-1}, z^{-2}, \cdots$ basis, while $n_p=1, \lvert p_0 \rvert <1$ corresponds to the Laguerre basis.



\noindent In H-OLOE the model is parameterized as 
\begin{equation}
G(q)=\frac{\sum_{k=1}^n \nu_k^TV_k(q)}{1+\sum_{k=1}^n \mu_k^TV_k(q)}=\frac{B(q)/A_p(q)}{A(q)/A_p(q)} 
\label{eq_G}
\end{equation}
\noindent  where $n \in \mathbb{N}$ and $n=n_a/n_p$. The monic polynomial $A_p(q)$ that we call the characteristic polynomial, whose roots are the basis poles, is defined as 
\begin{equation}
A_p(q)=\prod_{k=0}^{n_p-1}(q-p_k)^n
\label{eq_Ap}
\end{equation}
\noindent According to (\cite{r23}) and (\cite{r22}), the regressor of H-OLOE is given by
\begin{multline}
 {\phi^T_{\scalebox{0.6}{${HOLOE}$}}(t)=} \\
 {[-V_1^T(q)\hat{y}_{\scalebox{0.6}{$OE$}}(t+1), -V_2^T(q)\hat{y}_{\scalebox{0.6}{$OE$}}(t+1), \cdots  } \\
 {V_1^T(q)u(t+1), V_2^T(q)u(t+1), \cdots ] }
 \label{eq_reg_HOLOE}
 \end{multline}
  \noindent no algebraic loop occurs with such expression, since the function $V_k$ are strictly proper. The estimated parameters vector is
 \begin{equation}
\theta^T_{\scalebox{0.6}{${HOLOE}$}}(t)=[\mu_1^T \; \mu_2^T \; \cdots  \nu_1^T \; \nu_2^T \; \cdots ]
 \label{eq_param_vec_HOLOE}
 \end{equation}
  \noindent The prediction error is given by
 \begin{equation}
 \varepsilon_{\scalebox{0.6}{$OE$}}(t+1)=y(t+1)-\hat{y}_{\scalebox{0.6}{$OE$}}(t+1)
  \label{eq_pred_err_HOLOE}
 \end{equation}
  \noindent with $\hat{y}_{\scalebox{0.6}{$OE$}}(t+1)=\theta_{\scalebox{0.6}{${HOLOE}$}}^T \phi_{\scalebox{0.6}{${HOLOE}$}}(t)$. The parameters estimation is again carried out with the PALA.  And from the theory of equivalent prediction error, the expression of $Q(q,\theta)$ is now
 \begin{equation}
 \phi_{\scalebox{0.6}{${HOLOE}$}}(t)=-Q(q,\theta)\frac{\partial\varepsilon_{\scalebox{0.6}{$OE$}}(t+1)}{\partial\theta_{\scalebox{0.6}{${HOLOE}$}}(t)}
 \label{eq_Q_HOLOE1}
 \end{equation}
 \noindent Set \[\mathcal{A}(q, \theta_{\scalebox{0.6}{$\mathit{HOLOE}$}})=1+\sum_{k=1}^n \mu_k^TV_k(q)=\frac{A(q,\theta)}{A_p(q)}\]
 \noindent Combining \eqref{eq_reg_HOLOE}, \eqref{eq_param_vec_HOLOE}, \eqref{eq_pred_err_HOLOE}, \eqref{eq_Q_HOLOE1} yields
 \begin{equation}
 Q(q,\theta_{\scalebox{0.6}{$\mathit{HOLOE}$}})=\mathcal{A}(q, \theta_{\scalebox{0.6}{$\mathit{HOLOE}$}})
 \label{eq_Q_HOLOE}
 \end{equation}
 \noindent And the equivalent prediction error becomes
\begin{multline}
{\varepsilon_{E}(t+1,\theta_{\scalebox{0.6}{$\mathit{HOLOE}$}})=} \\
\mathcal{A}(q, \theta_{\scalebox{0.6}{$\mathit{HOLOE}$}})\varepsilon_{\scalebox{0.6}{$OE$}}(t+1,\theta_{\scalebox{0.6}{$\mathit{HOLOE}$}})+(1-\mathcal{A}(q, \theta_{\scalebox{0.6}{$\mathit{HOLOE}$}}))v(t+1)
\label{eq_eq_pred_err_HOLOE}
\end{multline}
\noindent yielding
\begin{multline}
\varepsilon_{E}(t,\theta_{\scalebox{0.6}{$\mathit{HOLOE}$}})=\\
\mathcal{A}(q, \theta_{\scalebox{0.6}{$\mathit{HOLOE}$}})\left(G_0(q)-G(q,\theta_{\scalebox{0.6}{$\mathit{HOLOE}$}})\right)u(t)+v(t)
\label{eq_eq_pred_err_HOLOE_direct}
\end{multline}
\noindent Consequently, if the true system is in the model set, the criterion $J_{HOLOE}$ minimized by H-OLOE is
\begin{equation}
J_{HOLOE}=\bar{\mathbf{E}}[\varepsilon_{E}^2(\theta_{\scalebox{0.6}{$\mathit{HOLOE}$}})]
\label{critere_HOLOE}
\end{equation}
\noindent with $\varepsilon_{E}(t,\theta_{\scalebox{0.6}{$\mathit{HOLOE}$}})$ given by \eqref{eq_eq_pred_err_HOLOE_direct}. From Lemma 1, the asymptotic estimated parameters vector of H-OLOE denoted as $\theta^*_{\scalebox{0.6}{${HOLOE}$}}$ also satisfies the relation
\begin{equation}
\bar{\mathbf{E}}[\varepsilon_{E}(t+1,\theta^*_{\scalebox{0.6}{${HOLOE}$}})\phi_{\scalebox{0.6}{${HOLOE}$}}(t,\theta^*_{\scalebox{0.6}{${HOLOE}$}})]=0
\label{eq_stationar3_HOLOE}
\end{equation}
\noindent even if the true system does not lie in the model set. But owing to \eqref{eq_eq_pred_err_HOLOE_direct} one also has  
\begin{equation}
\varepsilon_{E}(t,\theta_{\scalebox{0.6}{$\mathit{HOLOE}$}})=\varepsilon_{LSf}(t,\theta_{\scalebox{0.6}{$\mathit{HOLOE}$}})+(1-\mathcal{A}(q, \theta_{\scalebox{0.6}{$\mathit{HOLOE}$}}))v(t)
\label{}
\end{equation}
\noindent with 
 \begin{multline}
  {\varepsilon_{LSf}(t,\theta_{\scalebox{0.6}{$\mathit{HOLOE}$}})=} \\
  \scalebox{0.92}{$\displaystyle
    \mathcal{A}(q, \theta_{\scalebox{0.6}{$\mathit{HOLOE}$}}))
    \left(G_0(q)-G(q,\theta_{\scalebox{0.6}{$\mathit{HOLOE}$}})\right)u(t)
    +\mathcal{A}(q, \theta_{\scalebox{0.6}{$\mathit{HOLOE}$}})v(t)
  $} \\
  {=\frac{q^{n_a}}{A_p(q)}
  \varepsilon_{LS}(t,\theta_{\scalebox{0.6}{$HOLOE$}})}
  \label{eq_eq_errf}
\end{multline}
\noindent which is the equation error \eqref{eq1f} filtered by the AR system $\frac{q^{n_a}}{A_p(q)}$.Therefore the stationary condition \eqref{eq_stationar3_HOLOE} can be written
\begin{equation}
\bar{\mathbf{E}}[\varepsilon_{LSf}(t+1,\theta^*_{\scalebox{0.6}{${HOLOE}$}})\phi_{\scalebox{0.6}{${HOLOE}$}}(t,\theta^*_{\scalebox{0.6}{${HOLOE}$}})]=0
\label{eq_stationar4_HOLOEa}
\end{equation}
\noindent Now set 
\begin{equation}
 {  \phi^T_f(t)=\left[ -\hat{y}_{\scalebox{0.6}{$OEf$}}(t) \; -\hat{y}_{\scalebox{0.6}{$OEf$}}(t-1) \cdots  u_f(t) \; u_f(t-1) \cdots  \right]  } 
 \label{eq_phif}
  \end{equation}
\noindent such that $\hat{y}_{\scalebox{0.6}{$OEf$}}(t)=\frac{q^{n_a}}{A_p(q)}\hat{y}_{\scalebox{0.6}{$OE$}}(t)$ and $u_f(t)=\frac{q^{n_a}}{A_p(q)}u(t)$. From \eqref{eq_G} one has
$\hat{y}_{\scalebox{0.6}{$OE$}}(t+1)=\theta_{\scalebox{0.6}{${HOLOE}$}}^T \phi_{\scalebox{0.6}{${HOLOE}$}}(t)=\theta^T\phi_f(t)$ where $\theta$ is given in \eqref{eq_phi}. And, considering again \eqref{eq_G}, there exists a linear relation between $\phi_f(t)$ and $\phi_{\scalebox{0.6}{${HOLOE}$}}(t)$ which satisfy
\begin{equation}
\phi_f(t)=M\phi_{\scalebox{0.6}{${HOLOE}$}}(t) 
\label{eq_lin_phis}
\end{equation}
\noindent where $M$ is a square inversible matrix whose all elements belong to $\mathbb{R}$. Thus from \eqref{eq_stationar4_HOLOEa} and \eqref{eq_lin_phis}, the stationary condition of H-OLOE \eqref{eq_stationar3_HOLOE} is finally equivalent to
\begin{equation}
\bar{\mathbf{E}}[\varepsilon_{LSf}(t+1,\theta^*_{\scalebox{0.6}{${HOLOE}$}})\phi_{f}(t,\theta^*_{\scalebox{0.6}{${HOLOE}$}})]=0
\label{eq_stationar4_HOLOE}
\end{equation}
\noindent This is nothing other than the stationary condition of the SIV algorithm applied to input-output data filtered by the AR filter $\frac{q^{n_a}}{A_p(q)}$ (as in one iteration of SRIV). Denoting by $\theta^*_{\mathit{SIVf}}$ the corresponding asymptotic parameter vector, one therefore obtains
\begin{equation}
G(q,\theta^*_{\mathit{SIVf}})=G(q,\theta^*_{\scalebox{0.6}{${HOLOE}$}})
\label{eq_SIVf_HOLOE}
\end{equation}
\noindent regardless of whether the true system lies in the model set. The asymptotic models obtained by SIV with filtering previously input-output data by $\frac{q^{n_a}}{A_p(q)}$ is the same as the asymptotic model obtained with H-OLOE, where the basis poles of GBOFs are the roots of $A_p(q)$.
\begin{property}
Consider the asymptotic parameter vectors
$\theta^*_{\scalebox{0.6}{$\mathit{SIVf}$}}$ and
$\theta^*_{\scalebox{0.6}{$\mathit{HOLOE}$}}$
associated with the SIV algorithm using data
filtering by $\frac{q^{n_a}}{A_p(q)}$ and the
H-OLOE algorithm, respectively, where the
poles of the GBOFs are given by the roots
of $A_p(q)$. Assume that both identified models have the
same order $n_a$. Then, regardless of whether
the true system belongs to the model set,
the following equality holds:
\[
G(q,\theta^*_{\scalebox{0.6}{$\mathit{SIVf}$}}) =
G(q,\theta^*_{\scalebox{0.6}{$\mathit{HOLOE}$}}).
\]

Moreover, if the true system belongs to the
model set, both algorithms minimize the criterion $J_{2}$
defined by
\begin{multline}
{J_{2}=} \\
{\frac{1}{2\pi}
\int_{-\pi}^{\pi}
\left|
\frac{A(e^{i\omega},\theta)}{A_p(e^{i\omega})}
\right|^2
\left|
G_0(e^{i\omega})-G(e^{i\omega},\theta)
\right|^2
\Phi_u(\omega)\,d\omega}
\label{limit_mod_SIVf1}
\end{multline}
\label{prop2}
\end{property}
\noindent For the same reasons as in (\cite{r22}), the criterion \eqref{limit_mod_SIVf1} remains almost exact if the true system is no longer in the model set, provided that the bias is small in a given sense precisely defined in (\cite{r22}).

\subsection{Simulation example}

\begin{figure}
\begin{center}
\includegraphics[ width=3.4in, keepaspectratio ]{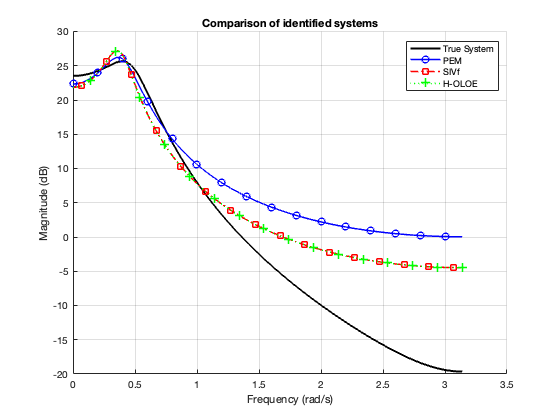}
\caption{Comparison of the limit models of SIV with filtered data by an AR filter (with poles equal to 0.5), HOLOE with predictor basis poles equal to 0.5 and PEM}
\label{fig2}
\end{center}
\end{figure}

The same system as in the simulation of Section 3.3 is now identified using SIV, but with all input-output data filtered by $\frac{q^{n_a}}{A_p(q)}$, where $n_a=2$ and all the filter poles are set to 0.5. The resulting model is compared with that obtained using H-OLOE (also with $n_a=2$), where all the basis poles of the GBOFs parameterizing the predictor are set to 0.5. It turns out that these two models are again identical, with a better fit at low frequencies than in the simulation of Section 3.3 (Fig.~1), but still differ from the model identified by PEM (see Fig.~\ref{fig2}).

\section{Analysis of iterative IV and PLR algorithms}

\subsection{The criterion minimized by I-H-OLOE and SRIV}

In (\cite{r23}), a procedure was also proposed that consists of running the H-OLOE algorithm several times, with the basis poles of the GBOFs at a given iteration chosen as the poles of the model obtained at the end of the previous H-OLOE identification. This iterative procedure was called I-H-OLOE. The term "iteration" should not be confused with the update of H-OLOE at each sampling instant.
Using the superscript $(l)$ to denote the $l$-th iteration, I-H-OLOE therefore consists of setting $A_p^{(l)}(q)=A^{(l-1)}(q)$. Similarly, the SRIV algorithm (\cite{r13}) consists of iteratively applying the SIV scheme to input-output data filtered by $\frac{1}{A_p(q)}$, such that one also has $A_p^{(l)}(q)=A^{(l-1)}(q)$.  It follows that, if I-H-OLOE and SRIV converge, one has
\begin{equation}
A^{(l)}- A_p^{(l)} \longrightarrow 0
\label{eq_asympt_A}
\end{equation}
\noindent  as $l \longrightarrow \infty $. One can now state the main result of this article: \\

\begin{theo}
Consider the asymptotic parameter vectors $\theta^*_{\scalebox{0.6}{$\mathit{SRIV}$}}$ and $\theta^*_{\scalebox{0.6}{$\mathit{IHOLOE}$}}$ associated with the SRIV and I-H-OLOE algorithms, respectively, and assume that both algorithms use models of the same order $n_a$. If both iterative algorithms converge, i.e., if $A^{(l)}-A_p^{(l)} \longrightarrow 0$ as the iteration number $l \longrightarrow \infty$, then, regardless of whether the true system belongs to the model set, both algorithms asymptotically minimize the criterion $J_{\scalebox{0.6}{$\mathit{3}$}}$, defined by
\begin{multline}
J_{3}=
\frac{1}{2\pi}
\int_{-\pi}^{+\pi}
\left|G_0(e^{i\omega})-G(e^{i\omega},\theta)\right|^2
\Phi_u(\omega)\,d\omega
\label{theo3_eq1}
\end{multline}
\end{theo}

\noindent Proof:
If $A^{(l)}-A_p^{(l)} \longrightarrow 0$ as $l \longrightarrow \infty$, then, from \eqref{eq_Q_HOLOE}, the transfer operator
$Q^{(l)}(q,\theta_{\scalebox{0.6}{$\mathit{HOLOE}$}})$ at iteration $l$ satisfies $Q^{(l)}(q,\theta_{\scalebox{0.6}{$\mathit{HOLOE}$}}) \longrightarrow 1$ as $l \longrightarrow \infty$.
Consequently, the equivalent prediction error of H-OLOE at iteration $l$, denoted by
$\varepsilon_E^{(l)}(t,\theta_{\scalebox{0.6}{$\mathit{HOLOE}$}})$ and defined in \eqref{eq_eq_pred_err_HOLOE_direct}, satisfies
$\varepsilon_E^{(l)}(t,\theta_{\scalebox{0.6}{$\mathit{HOLOE}$}}) \longrightarrow \varepsilon_{\scalebox{0.6}{$OE$}}(t,\theta_{\scalebox{0.6}{$\mathit{HOLOE}$}})=
\left(G_0(q)-G(q,\theta_{\scalebox{0.6}{$\mathit{HOLOE}$}})\right)u(t)+v(t)$. 
Similarly, the H-OLOE regressor at iteration $l$, denoted here by $\phi^{(l)}_{\scalebox{0.6}{$HOLOE$}}(t)$ and defined in \eqref{eq_reg_HOLOE}, satisfies $\phi^{(l)}_{\scalebox{0.6}{$HOLOE$}}(t) \longrightarrow-\frac{\partial
\varepsilon_{\scalebox{0.6}{$OE$}}(t+1,\theta_{\scalebox{0.6}{$\mathit{HOLOE}$}})} {\partial\theta_{\scalebox{0.6}{$\mathit{HOLOE}$}}}$ as $l \longrightarrow \infty$, according to \eqref{eq_Q_HOLOE1}. Therefore, as $l \longrightarrow \infty$,
the generic stationarity condition \eqref{eq_stationar3_PLR} of Lemma 1, which remains valid
even though the true system does not belong to the model set, becomes \\
$\bar{\mathbf{E}}\left[
\varepsilon_{\scalebox{0.6}{$OE$}}
(t+1,\theta^*_{\scalebox{0.6}{$\mathit{IHOLOE}$}})
\left.
\frac{\partial
\varepsilon_{\scalebox{0.6}{$OE$}}(t+1,\theta)}
{\partial\theta}
\right|_{\theta=
\theta^*_{\scalebox{0.6}{$\mathit{IHOLOE}$}}}
\right]=0$.
This is precisely the stationarity condition of PEM given in \eqref{eq_stationar_PEM}. It follows that
$G(q,\theta^*_{\scalebox{0.6}{$IHOLOE$}})=G(q,\theta^*_{\scalebox{0.6}{$PEM$}})$, and similarly, $G(q, \theta^*_{\scalebox{0.6}{$SRIV$}})=G(q,\theta^*_{\scalebox{0.6}{$PEM$}})$.
The latter equality follows from Property \ref{prop2}, which states that
$G(q,\theta^*_{\scalebox{0.6}{$HOLOE$}})=G(q,\theta^*_{\scalebox{0.6}{$SIVf$}})$
provided that the model order $n_a$ is the same in both cases. Here, $\theta^*_{\scalebox{0.6}{$SIVf$}}$ is obtained using SIV with data filtering by $\frac{q^{n_a}}{A_p(q)}$, where $A_p$ is also
the characteristic polynomial of the GBOFs used in H-OLOE. Consequently,
$\theta^*_{\scalebox{0.6}{$\mathit{SRIV}$}}= \operatorname*{arg\,min}_{\theta}\bar{\mathbf{E}}[
\varepsilon^2_{\scalebox{0.6}{$\mathit{OE}$}}(t,\theta)]$ and $\theta^*_{\scalebox{0.6}{$\mathit{IHOLOE}$}} = \operatorname*{arg\,min}_{\theta}
\bar{\mathbf{E}}[\varepsilon^2_{\scalebox{0.6}{$\mathit{OE}$}}(t,\theta)]$,
regardless of whether the true system belongs to the model set. Expressing these conditions in the frequency
domain yields \eqref{theo3_eq1}. $\Box$

\noindent  I-H-OLOE and SRIV both aim indirectly at minimizing asymptotically the variance $\bar{\mathbf{E}}[\varepsilon^2_{\scalebox{0.6}{$\mathit{OE}$}}(t,\theta)]$ of the prediction error $\varepsilon_{\scalebox{0.6}{$\mathit{OE}$}}(t,\theta)$ given in \eqref{eq1c} and can therefore be considered as indirect ways of achieving the goal of PEM regardless of whether the true system lies in the model set. Furthermore, if the true system effectively is in the model set and if $\{v(t)\}$ is Gaussian white noise, SRIV and I-H-OLOE become, like PEM, statistically efficient (as demonstrated by other means in (\cite{r26}) for SRIV and (\cite{r23}) for I-H-OLOE), due to the asymptotic minimization of the prediction error variance.

\subsection{Simulation example}

\begin{figure}
\begin{center}
\includegraphics[ width=3.4in, keepaspectratio ]{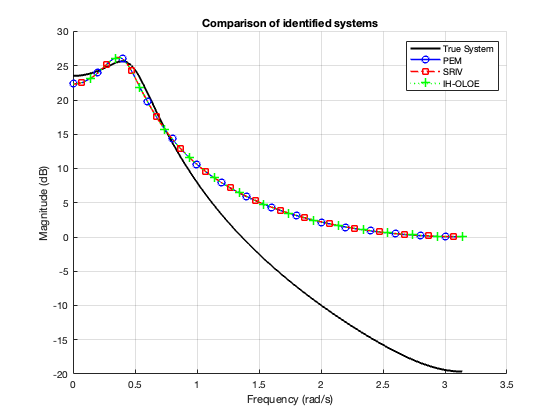}
\caption{Comparison of the limit models of SRIV, I-H-OLOE and PEM}
\label{fig3}
\end{center}
\end{figure}

We consider again the same system, identified under the same conditions, and now compare SRIV, I-H-OLOE, and PEM. The three Bode diagrams of the identified systems are visually indistinguishable, confirming that SRIV and I-H-OLOE, even in the presence of bias, minimize a criterion that is the prediction error variance (see Fig.~\ref{fig3}).

\section{Concluding remarks}

This paper has brought to light the connection between the criteria associated with Instrumental Variable (IV) and Pseudo Linear Regression (PLR) schemes for the identification of discrete-time output error models. Upon convergence, the original IV algorithm with an auxiliary model (SIV) minimizes the same criterion as the Output Error version of PLR (called the Landau scheme), regardless of whether the true system lies within the model set. This observation remains valid when the input-output data used by SIV are filtered through an AR filter whose poles coincide with the basis poles of the PLR scheme (H-OLOE) based on Generalized Orthonormal Basis Functions. Finally, it has been established that the iterative IV scheme (SRIV) and the iterative PLR scheme (I-H-OLOE) both aim to minimize the prediction error variance. Consequently, these two iterative algorithms can be interpreted as indirect Prediction Error Methods (PEMs). This finding offers a new perspective on these three classes of algorithms.

\end{document}